\documentclass[
 reprint,
 amsmath,amssymb,
 aps]{revtex4-2}

\usepackage{graphicx}
\usepackage{dcolumn}
\usepackage{bm}
\usepackage{hyperref}
\usepackage{xcolor}

\newcommand{\be}{\begin{equation}}
\newcommand{\ee}{\end{equation}}
\newcommand{\bea}{\begin{eqnarray}}
\newcommand{\eea}{\end{eqnarray}}
\newcommand{\vo}{\mathcal{V}}

\begin{document}

\preprint{APS/123-QED}

\title{The Dark Universe after Reheating in String Inflation}

\author{Michele Cicoli}
\email{michele.cicoli@unibo.it}
\affiliation{Dipartimento di Fisica e Astronomia, Universit\`a di Bologna, via Irnerio 46, 40126 Bologna, Italy \\
and INFN, Sezione di Bologna, viale Berti Pichat 6/2, 40127 Bologna, Italy
}
\author{Kuver Sinha and Robert Wiley Deal}
\email{kuver.sinha@ou.edu, rwileydeal@ou.edu}
\affiliation{Department of Physics and Astronomy, University of Oklahoma, Norman, OK 73019, USA}

\date{\today}

\begin{abstract}
We study the production of dark matter and dark radiation after reheating in string inflation models where the Calabi-Yau has a fibred structure and the visible sector lives on D3 branes. We show how the interplay between different physical constraints from inflation, reheating, supersymmetry breaking and dark radiation, leads to distinct predictions for the nature of dark matter.
\end{abstract}

\maketitle

\tableofcontents

\section{Introduction}
\label{sec:level1}

Recent advances in string phenomenology in type IIB compactifications have progressed in two complementary directions: 

($i$) Constructing specific compactifications that are phenomenologically  promising: typically, these constructions  incorporate moduli stabilization, have chiral matter, are  broadly able to accommodate the gauge groups and matter content of the Minimal Supersymmetric Standard Model (MSSM) \cite{Cicoli:2011qg, Cicoli:2012vw, Cicoli:2013mpa, Cicoli:2013cha, Cicoli:2021dhg, Cvetic:2022fnv}, and have at least a somewhat well-defined inflationary and reheating sector \cite{Cicoli:2016xae, Cicoli:2017shd, Cicoli:2017axo}; 

($ii$) Proceeding along statistical lines, by drawing statistical conclusions about the distribution of important phenomenological quantities like the scale of supersymmetry breaking and the axion decay constant, from the ensemble of type IIB flux vacua \cite{Douglas:2012bu, Denef:2006ad, Denef:2004ze, Broeckel:2020fdz, Broeckel:2021dpz, Broeckel:2021uty, Demirtas:2021nlu, Demirtas:2020dbm}. 

Amongst various classes of models in the first direction, type IIB Large Volume flux compactifications are particularly well-developed. Two main inflationary scenarios emerge in this context: K\"ahler moduli inflation (KMI) \cite{Conlon:2005jm, Bond:2006nc} and Fibre Inflation (FI) \cite{Cicoli:2008gp, Broy:2015zba, Cicoli:2016chb}. KMI is a small-field model where inflation is driven by a blow-up mode with a non-perturbative scalar potential. The Hubble scale during inflation is relatively low, $H_I\sim 5\times 10^8$ GeV, and the tensor-to-scalar ratio is unobservable $r\simeq 10^{-10}$. On the other hand, FI is a large field model characterized by $H_I\simeq 5\times 10^{13}$ GeV and $r\simeq 0.007$ \cite{Cicoli:2020bao}, where the inflaton is a fibration bulk modulus with a perturbative scalar potential. 

In these constructions the visible sector can live on either D7-branes wrapping 4-cycles in the geometric regime, or D3-branes at singularities. In the first case, the soft terms are around the gravitino mass, $M_{1/2}\sim m_0 \sim m_{3/2}$, while in the second case the visible sector can be sequestered from the sources of supersymmetry breaking in the bulk, resulting in soft terms which can be hierarchically smaller than the gravitino mass\cite{Blumenhagen:2009gk, Aparicio:2014wxa}. Two limits can arise: a so-called \emph{local limit} with a split-SUSY spectrum featuring $M_{1/2}\ll m_0 \ll m_{3/2}$, and an \emph{ultralocal limit} with a more standard MSSM-like spectrum with $M_{1/2}\sim m_0 \ll m_{3/2}$.

Reheating via the decay of the modulus with the smallest decay width has already been studied in both KMI and FI for several D-brane configurations which can realize an MSSM-like sector together with additional hidden sectors.
In particular, reheating for KMI in the simplest Swiss-cheese LVS models has been studied in \cite{Cicoli:2010yj, Cicoli:2012aq, Higaki:2012ar} for D3-branes in the ultralocal limit, in \cite{Cicoli:2015bpq} for D3-branes in the local limit, and in \cite{Cicoli:2010ha, Hebecker:2014gka} for D7-branes. On the other hand, reheating for FI with the MSSM on D3-branes has been studied in \cite{Angus:2014bia}, and in \cite{Cicoli:2018cgu} for the D7-brane case. Each of these references analyzed in detail the constraints arising from the requirement to avoid an excessive production of ultra-light bulk axions that behave as dark radiation. Moreover \cite{Allahverdi:2013noa, Allahverdi:2014ppa, Aparicio:2015sda, Aparicio:2016qqb} studied the implications for non-thermal neutralino dark matter for KMI with the MSSM on D3-branes in the ultralocal limit. The constraints on the nature of dark matter for KMI with the visible sector on D7-branes has instead been analyzed in \cite{Allahverdi:2020uax} for superheavy WIMPs and in \cite{Hebecker:2022fcx} for the QCD axion realized as a closed string mode. 

The largest production of dark radiation from the decay of the lightest modulus has been found in \cite{Angus:2014bia} for FI with the MSSM on D3-branes. This result relies on a particular expression of the moduli-dependence of the Giudice-Masiero contribution to the K\"ahler potential which determines an effective decoupling of the lightest modulus from the Higgs degrees of freedom. In this paper we will revisit this result by considering a more general moduli-dependence of the Giudice-Masiero term that allows to considerably reduce the production of dark radiation. In doing so, we shall follow the results of \cite{Aparicio:2008wh} which constrained the form of the K\"ahler potential by analogy with explicit toroidal computations. 

We will then study the associated production of dark matter after reheating in FI. We will find that WIMPs are always overproduced, requiring a mechanism of R-parity breaking to make them unstable. In this case, a very promising dark matter candidate is instead the QCD axion realized as the phase of a charged open string field. The axion decay constant is around $f_{\rm QCD}\simeq 5\times 10^{10}$ GeV, which can avoid isocurvature bounds and lead to dark matter in a rather natural way. We will extend our analysis also to fibred Calabi-Yau compactifications with KMI and the MSSM on D3-branes. The best case scenario to avoid dark radiation overproduction with the minimal tuning of the coefficient of the Giudice-Masiero interaction among the moduli and the Higges, is the ultralocal limit. In this case, we will find that dark matter can be TeV-scale non-thermal neutralinos which are produced from the inflaton decay and then undergo annihilation. 

Our analysis shows how UV correlations among different observables in string compactifications, like inflation, supersymmetry breaking and reheating, is very powerful to obtain clear predictions for the nature of dark matter. Moreover, we will see that avoiding dark radiation overproduction severely constrains the moduli-dependence of the K\"ahler metric for matter fields. 

We will also point out that, contrary to what was claimed in previous studies \cite{Cicoli:2012aq}, the moduli decay to open string axions can compete with the decay to bulk axions, and so lead to an additional source of dark radiation. 

This paper is organized as follows. In Section \ref{nonthermalDM} and Section \ref{CYmodels}, we provide a brief review of non-thermal dark matter and fibred LVS models, respectively. In Section \ref{general}, we first derive the moduli couplings to closed and open string axions, as well as with other visible sector fields. We then derive the associated predictions for dark radiation and the reheat temperature. In Sections \ref{case1a} and \ref{case2a}, we study non-thermal dark matter candidates in KMI and FI, respectively, before presenting our conclusions in Section \ref{Concl}. 

\section{Non-thermal dark matter}
\label{nonthermalDM}

Before focusing on string models, we first briefly review non-thermal dark matter (DM) produced by heavy scalar decay \cite{Kane:2015jia, Allahverdi:2020bys}. The lightest modulus, $\phi$, dominates the energy density of the universe from the onset of its oscillations until its decay. The decay then produces a significant amount of entropy, diluting any previously existing DM particles to a negligible abundance. Depending on the reheat temperature of $\phi$, $T_{\text{rh}}$, the DM particles produced by $\phi$ decay may fall into a few different scenarios depending on the DM freeze-out temperature $T_{\rm f}\simeq m_{\rm DM}/20$ \cite{Allahverdi:2013noa}.
\paragraph{$T_{\text{rh}} \gtrsim T_{\text{f}}$}
In this case, the DM may equilibrate. Due to the DM-DM annihilations, the DM abundance is simply given by that of the thermal scenario.

\paragraph{$T_{\text{rh}} \lesssim T_{\text{f}}$}
Here, we have two sub-cases depending on the efficiency of the DM annihilations. More precisely, defining the critical abundance of the DM, $Y_{\text{DM}}^c$, as the attractor of the relevant Boltzmann equation, we have 
\begin{align}
Y_{\text{DM}}^c  \simeq  \frac{H}{\langle \sigma_{\text{ann}} v \rangle s} \Big|_{T_{\text{rh}}}.
\end{align}

If the produced DM abundance is larger than the critical abundance, i.e. $Y_{\text{DM}} > Y_{\text{DM}}^c$, some annihilations still occur until the abundance reaches $Y_{\text{DM}}^c$, at which point the DM becomes too dilute to annihilate efficiently. We refer to this case as the \textit{annihilation scenario.} In the annihilation scenario, the final DM abundance can be estimated by 
\begin{align}
\label{eq:annihilationDMAbundance}
Y_{\text{DM}}^c \simeq \left(\frac{ n_{\text{DM}} }{s} \right)_{\text{obs}}  \frac{ \langle \sigma_{\text{ann}} v \rangle_{\text{f}}^{\text{th}} }{\langle \sigma_{\text{ann}} v \rangle_{\text{f}}  }\left( \frac{ T_{\text{f}} }{T_{\text{rh}} }  \right)
\end{align}
where $\langle \sigma_{\text{ann}} v \rangle_{\text{f}}^{\text{th}} \simeq 3 \times 10^{-26} \text{ cm}^3 s^{-1}$ is the value needed in the thermal case \cite{Fermi-LAT:2016uux} to match the observed DM abundance
\begin{align}
\label{dmObservedAbundance}
\left( \frac{ n_{\text{DM}} }{s} \right)_{\text{obs}} 
\simeq 5 \times 10^{-10}\left( \frac{1 \text{ GeV} }{m_{\text{DM}} } \right).
\end{align}
The DM abundance in the annihilation scenario is thus enhanced by $T_{\text{f}} / T_{\text{rh}}$ in comparison to the thermal case, and can accommodate thermally underproduced DM candidates, such as wino-like and higgsino-like DM. Since $T_{\text{rh}} \lesssim T_f \sim m_{\text{DM}} / 20$, we must have $\langle \sigma_{\text{ann}} v \rangle_f^{\text{th}} < \langle \sigma_{\text{ann}} v \rangle_f$ to reproduce the observed DM abundance.

If the produced DM abundance is lower than the critical abundance, i.e. $Y_{\text{DM}} < Y_{\text{DM}}^c$, then annihilations cannot occur significantly. We refer to this case as the \textit{branching scenario.} The late-time DM abundance is given simply by the fraction of the light modulus abundance, $Y_\phi$, that decays to DM
\begin{align}
Y_{\text{DM}} = Y_\phi \, \text{Br}_{\text{DM}}
.
\label{YDMbranching}
\end{align}
If the branching ratio to DM is sufficiently low, the branching scenario can accommodate thermally overproduced DM, such as bino-like DM.

\section{Fibred Calabi-Yau Models}
\label{CYmodels}

Fibred Calabi-Yau manifolds have been shown to be very promising for cosmological and phenomenological applications. The simplest setup involves three K\"ahler moduli. To settle notation for the rest of the paper, we briefly review this class of compactifications.

The first K\"ahler modulus we consider is $T_1 = \tau_1 + {\rm i} c_1$. The scalar $\tau_1$ parametrizes the dimensionless volume in string units of a K3 or $T^4$ fibre over a $\mathbb{P}^1$ base. The axionic component $c_1$ arises from the reduction of the 10D RR form $C_4$ on the K3 or $T^4$ divisor. The scalar component of the second modulus $T_2 = \tau_2 + {\rm i} c_2$ controls the volume of the base of the fibration, while $c_2$ is the associated axion. The scalar component of the third modulus $T_3 = \tau_3 + {\rm i} c_3$ parametrizes the size of a blow-up mode required to stabilize the volume at leading order, and $c_3$ is an additional closed string axion. The volume takes the form (for explicit Calabi-Yau models with this volume form see \cite{Cicoli:2011it,Cicoli:2016xae,Cicoli:2017axo})
\be
\mathcal{V}  = \sqrt{\tau_1}\tau_2 - \tau^{3/2}_3\,.
\label{voform}
\ee
The K\"ahler potential and the superpotential read
\bea
K &=& K_{\rm tree} + K_{\alpha'} + K_{g_s} \nonumber \\
W &=& W_0 + A_1 e^{-\mathfrak{a}_1 T_1} + A_2 e^{-\mathfrak{a}_2 T_2} + A_3 e^{-\mathfrak{a}_3 T_3} .
\eea
The tree-level K\"ahler potential is $K_{\rm tree} = - 2 \ln{\mathcal{V}}$ (setting $m_P=1$) and it enjoys a well-known no-scale property which makes all K\"ahler moduli flat at semiclassical level. At this order of approximation, background 3-form fluxes freeze the complex structure moduli and the dilaton whose VEV sets the string coupling $g_s$ which we assume to be in the weak coupling regime, i.e. $g_s\lesssim\mathcal{O}(0.1)$, to trust perturbation theory. The leading order $\alpha'^3$ correction is given by $K_{\alpha'} = -\frac{\xi}{g^{3/2}_s\mathcal{V}}$, where $\xi$ is the Calabi-Yau Euler number, $\xi\sim\mathcal{O}(1)$, while for string loop corrections to $K$, given by $K_{g_s}$, we refer to \cite{Cicoli:2018cgu}. The superpotential contains instead the tree-level flux-generated contribution $W_0$ (which is a tunable constant after complex structure and dilaton stabilization), and non-perturbative corrections for each of the three K\"ahler moduli, where the $A_i$'s are expected to be $\mathcal{O}(1)$ constants while $\mathfrak{a}_i=2\pi/N_i$ with $N_i \in \mathbb{N}$.

In the large volume limit where the effective field theory is under control, $\tau_2 \gtrsim \tau_1\gg \tau_3 >1$, the leading order contribution to the potential for the K\"ahler moduli arises from $K_{\alpha'}$ and the $T_3$-dependent non-perturbative correction to $W$. This potential fixes $\tau_3$, $c_3$ and the overall volume $\vo$ at
\be
\langle\tau_3\rangle\simeq g_s^{-1}\qquad \langle\vo\rangle\simeq W_0\,e^{\mathfrak{a}_3/g_s} \qquad \langle c_3\rangle= \pi/\mathfrak{a}_3.
\ee
The minimum is AdS and breaks supersymmetry spontaneously. There are several known mechanism to uplift it to Minkowski, among which anti-D3 branes \cite{Kachru:2003aw}, T-branes \cite{Cicoli:2015ylx}, or non-zero F-terms of the complex structure moduli \cite{Gallego:2017dvd}. At subleading order, $K_{g_s}$ (or higher order $\alpha'$ corrections \cite{Ciupke:2015msa,Cicoli:2016chb}), fix the direction orthogonal to the volume mode $u\equiv \tau_1/\tau_2$ at \cite{Cicoli:2018cgu}
\be
\langle u\rangle = \lambda\, g_s^2\,,
\ee
where $\lambda$ depends on the tunable coefficients of string loop corrections to $K$. Finally, tiny $T_1$- and $T_2$-dependent non-perturbative effects fix the two ultra-light axions $c_1$ and $c_2$ at $\langle c_i\rangle = \pi/\mathfrak{a}_i$ $\forall i=1,2$. 

The gravitino mass takes the form (restoring appropriate powers of $m_P$)
\be
m_{3/2}^2 \simeq \kappa \epsilon^2 m_P^2
\ee
where \cite{Burgess:2010bz}
\be
\epsilon \equiv W_0/\vo \ll 1\quad \text{and}\quad\kappa\equiv g_s/(8\pi)\ll 1
\ee
and the moduli mass spectrum is given by \cite{Cicoli:2010ha,Cicoli:2010yj,Cicoli:2017zbx}
\begin{align}
\label{eq:massSpectrum}
m_{\tau_3}^2    &\simeq  m_{c_3}^2 \simeq   \left(\ln \epsilon \right)^2  m_{3/2}^2 > m_{3/2}^2 \nonumber \\
m_\vo^2    &\simeq \left(\frac{\epsilon}{ g_s^{3/2}  W_0| \left( \ln \epsilon \right)^3|}\right) m_{\tau_3}^2 \ll m_{\tau_3}^2  \nonumber    \\
m_u^2    &\simeq   \left( \frac{ \epsilon^{1/3}
g_s^{5/6} |\ln \epsilon|}{W_0^{1/3} 
\sqrt{\lambda}}\right)m_\vo^2  < m_\vo^2 \nonumber    \\
m_{c_i}^2 &\simeq  \frac{\tau_i^3}{W_0}\,e^{-\mathfrak{a}_i\tau_i} m_{3/2}^2 \ll m_u^2\quad \forall i=1,2
\end{align}
The visible sector lives on D3-branes at singularities which can give rise to a scenario of sequestered SUSY breaking \cite{Blumenhagen:2009gk}. The common gaugino masses, $M_{1/2}$, can then be estimated using the results of \cite{Aparicio:2014wxa}
\be
M_{1/2} \simeq \frac{3\omega}{2}   \frac{m_{3/2}}{\mathcal{V}}\, \tau_3^{3/2} \sim \mathcal{O} \left(\frac{m_{3/2}}{\mathcal{V}} \left( \ln \mathcal{V} \right)^{3/2} \right)
\label{M12}
\ee
where $\omega$ is a tunable flux-dependent parameter. The $\mu$-term, if generated by a Giudice-Masiero contribution to the K\"ahler potential, is expected to be approximately of order $M_{1/2}$ since it turns out to be $\mu\simeq \tilde\omega \,M_{1/2}$ where $\tilde\omega$ is another tunable flux-dependent parameter. On the other hand, the value of the common slepton and squark mass, denoted as $m_0$, depends on the exact moduli-dependence of the K\"ahler metric for matter fields. The so-called ``local limit'' is defined as the case where the physical Yukawas are independent of the volume \textit{only to leading order} in $\mathcal{V}^{-1}$, while the ``ultralocal limit'' is defined to be the case where the physical Yukawas are independent of the volume \textit{to any order} in $\mathcal{V}^{-1}$ \cite{Aparicio:2014wxa}. Consequently, the local limit leads to a split SUSY scenario, whereas the ultralocal case can give a standard MSSM-like spectrum
\bea
&&\text{local limit}: \qquad\quad m_0 \simeq M_{1/2}\sqrt{\vo} \gg M_{1/2}\nonumber \\
&&\text{ultralocal limit}:\quad m_0 \simeq M_{1/2} .
\label{limits}
\eea

\section{Moduli decays and dark radiation} 
\label{general}

As argued above, reheating is given by the decay of the lightest modulus, which in the case of fibred Calabi-Yaus is $u$. This K\"ahler modulus can decay to visible sector fields on D3-branes at singularities, but also to the two ultra-light closed string axions $c_1$ and $c_2$ which behave as dark radiation. 

The decay to SUSY scalars is kinematically forbidden in the local limit, whereas it is mass-suppressed in the ultralocal limit. Similarly, the decay to SM fermions, Higgsinos and gauginos is chirality suppressed. Moreover, the decay to SM gauge bosons is loop suppressed since the gauge kinetic function for D3-branes is controlled by the dilaton. The only relevant decay rates are therefore to Higgses via a Giudice-Masiero interaction term in $K$, and to the closed string axions $c_1$ and $c_2$. We will also argue that, if the QCD axion is realized as the phase of a charged open string field, the lightest modulus decay into this mode should also be taken into account since it is not mass-suppressed. 

Assembling all the results, we then provide constraints on the parameter space coming from observed upper bounds on the effective number of neutrino species $N_{\rm eff}$.

\subsection{Canonical normalization}

We begin by generalizing the volume form (\ref{voform}) to  
\be
\vo =\tau_1^{\frac{n_1}{2}}\tau_2^{\frac{n_2}{2}}
\ee
with the implicit constraint that $n_1 + n_2 = 3$. Notice that we ignored the blow-up mode $\tau_3$ given that it does not play any relevant role in reheating. This volume corresponds to a K\"ahler potential of the form 
\begin{align}
\label{eq:modulusKahlerPotential}
\frac{K}{m_P^2} & = -n_1 \ln(T_1+\overline{T}_1 ) -  n_2  \ln(T_2+\overline{T}_2 ).
\end{align}
The K\"ahler metric, $K_{i \overline{\jmath}} \equiv \partial_i \partial_{\overline{\jmath}} K$, is then given by 
\begin{align}
K_{i\overline{\jmath}}    &= \frac{m_P^2}{4}   \begin{pmatrix}
\frac{n_1}{\tau_1^2} && 0 \\
0 && \frac{n_2}{\tau_2^2}
\end{pmatrix}.
\end{align}
As expected, since the K\"ahler potential is separable, we have already a diagonal K\"ahler metric.

The kinetic term in the Lagrangian
\begin{align}
\label{kineticLagrangianModuli}
\mathcal{L} &= K_{i \overline{\jmath}}    \partial_\mu T^i \partial^\mu \overline{T}^{\overline{\jmath}}    \supset \frac{m_P^2}{4}\frac{n_i}{\tau_i^2}  \partial_\mu \tau_i \partial^\mu \tau_i
\end{align}
can then be put into canonical form with the field redefinitions 
\begin{align}
\label{redefinitions}
\tau_i &= \exp \left(\sqrt{\frac{2}{n_i}} \frac{\phi_i}{m_P} \right)
\end{align}
where $\phi_i$ are the new fields with canonical kinetic terms.

There is, however, still a degeneracy - the moduli $\phi_i$ are, in general, not mass eigenstates. As we have seen in Section \ref{CYmodels}, the Calabi-Yau volume is fixed by the leading order dynamics, and so the mass eigenstates are the volume mode and its transverse direction. The volume mode is given by 
\begin{align}
\frac{\phi_\vo}{m_P} &\propto \ln \mathcal{V} 
= \frac12 \ln \left( \tau_1^{n_1} \tau_2^{n_2}\right).
\end{align}
Plugging in our field redefinitions, Eq.~(\ref{redefinitions}), into this expression and multiplying by an overall normalization constant, we have the volume mode in terms of $\phi_i$:
\begin{align}
\phi_\mathcal{V} &= \sqrt{ \frac{n_1}{n_1 + n_2}    }\,\phi_1 +  \sqrt{\frac{n_2}{n_1 + n_2}}\, \phi_2.
\end{align}
The transverse mode, $\phi_u$, can then be constructed simply by orthogonality:
\begin{align}
\phi_u &=   -  \sqrt{\frac{n_2}{n_1 + n_2}}\, \phi_1 + \sqrt{ \frac{n_1}{n_1 + n_2}}\, \phi_2.
\end{align}
We will also utilize the inverse of these transformations:
\begin{align}
\label{massEigenbasisTransformation}
\phi_1 &= \sqrt{\frac{n_1}{n_1 + n_2}}    \,\phi_{\mathcal{V}} -\sqrt{\frac{n_2}{n_1 + n_2} } \,\phi_u \nonumber \\    \phi_2   &= \sqrt{\frac{ n_2 }{n_1 + n_2} }\,\phi_{\mathcal{V}}+\sqrt{\frac{n_1}{n_1 + n_2}}\,  \phi_u.
\end{align}

\subsection{Decays to closed string axions}

We now discuss the moduli decays into closed string axions $c_i$. We start by returning to the kinetic term (\ref{kineticLagrangianModuli}), which also contains the terms 
\begin{align}
\label{kineticLagrangianALPs}
\mathcal{L} &= K_{i\overline{\jmath}} \partial_\mu T^i \partial^\mu \overline{T}^{\overline{\jmath}}
\supset \frac{m_P^2}{4}\frac{n_i}{\tau_i^2} \partial_\mu  c_i \partial^\mu c_i.
\end{align}
Applying the field redefinition (\ref{redefinitions}), expanding the exponential, and rescaling the axion fields by
\begin{align}
c_i &= \sqrt{\frac{2}{n_i}}\, \frac{a_i}{m_P}
\end{align}
we obtain canonical kinetic terms for the axion fields $a_i$ in addition to the interaction terms
\begin{align}
\mathcal{L}    &\supset - \sqrt{\frac{2}{n_i}}\, \frac{\phi_i}{m_P} \, \partial_\mu a_i\partial^\mu a_i.
\end{align}

Applying now the transformations into the moduli mass eigenbasis, Eq.~(\ref{massEigenbasisTransformation}), we arrive at 
\begin{align}
\mathcal{L}    &\supset  -\sqrt{ \frac{2}{3} }\frac{\phi_\vo}{m_P}\left( \partial_\mu a_1 \partial^\mu a_1+\partial_\mu a_2    \partial^\mu a_2    \right)      \nonumber    \\ &  -  \sqrt{\frac{2}{3}} \frac{\phi_u}{m_P}\left(\sqrt{  \frac{n_1}{n_2}}\, \partial_\mu a_2  \partial^\mu  a_2 - \sqrt{\frac{n_2}{n_1}}\, \partial_\mu a_1 \partial^\mu a_1 \right)
\end{align}
where we have explicitly used the constraint $n_1 + n_2 = 3$.

Focusing on the transverse mode, $\phi_u$, we can now write down the total decay width to closed string axions (considering both axions to be massless)
\begin{align}
\label{eq:alpDecayWidth}
\Gamma(\phi_u \rightarrow a a) &=    \frac{1}{48 \pi}  \left(\frac{ n_1^2 +n_2^2 }{n_1 n_2 }\right)    \frac{m_u^3}{m_P^2}\equiv c_{\rm hid}\, \Gamma_0
\end{align}
where we have defined
\begin{align}
c_{\rm hid} & \equiv\left(\frac{n_1^2+ n_2^2}{n_1 n_2} \right)
\end{align}
to be the coefficient of hidden sector decays.
We have also made the definition
\begin{align}
\Gamma_0 &\equiv\frac{1}{48 \pi} \frac{m_u^3}{m_P^2}
\end{align}
for future convenience.

\subsection{Decays to open string axions}
\label{OpenAxions}

Open string axions, which arise as the phase of charged matter fields acquiring a non-zero VEV, are more model-dependent than the closed string bulk axions we considered above. In \cite{Cicoli:2012aq}, the modulus decay to open string axions was considered and claimed to be negligible. Here, we demonstrate that this is not necessarily the case - both within our fibred LVS scenario and the minimal LVS scenario considered in \cite{Cicoli:2012aq}. We begin by considering open string axions within the context of minimal LVS with only one bulk K\"ahler modulus corresponding to the volume mode. For matter fields, collectively denoted as $C$, the relevant term in the K\"ahler potential is 
\begin{align}
\label{eq:matterKahlerPotentialMinimalLVS}
\frac{K}{m_P^2} & \supset \frac{C\overline{C}}{ T_b + \overline{T}_b}
\end{align}
where $T_b$ is the bulk modulus, corresponding to the choice $n_1 = 3$, $n_2 = 0$ in Eq.~(\ref{eq:modulusKahlerPotential}).
We can then write down the Lagrangian for the canonical moduli after using Eq.~(\ref{redefinitions})
\begin{align}
\label{eq:matterFieldLagrangian}
\mathcal{L}    &    \supset    \frac{m_P}{2\sqrt{6}}\,\phi_\vo \left( C \Box  \overline{C}  + \overline{C}  \Box C    \right).
\end{align}

The matter field can then be parameterized by $C = \rho \, \exp({\rm i} \, \theta), $ where $\rho$ is the radial component which acquires a VEV $\langle \rho \rangle \neq 0$ via D-term stabilization, while $\theta$ is the phase field taking the role of the axion. Once $\rho$ takes on its VEV, and after going to canonically normalized fields $\tilde{\rho}$ and $\tilde{\theta}$, defined as $\rho = \sqrt{\langle\tau_b\rangle}\tilde{\rho}/m_P$ and $\tilde{\theta} = \langle\tilde{\rho}\rangle\theta$,  Eq.~(\ref{eq:matterFieldLagrangian}) gives
\be
\mathcal{L}  \supset    -    \frac{1}{\sqrt{6}}\,\frac{\phi_\vo}{m_P} \, \partial_\mu \tilde\theta \,\partial^\mu \tilde\theta.
\label{open}
\ee
Notice that the axion decay constant $f_\theta$ is set by $\langle\tilde{\rho}\rangle$, i.e. $f_\theta=\langle\tilde{\rho}\rangle$. To put Eq.~(\ref{open}) in a more illuminating form, we can integrate by parts which gives us 
\begin{align}
\mathcal{L} & \supset \frac{1}{2\sqrt{6}\,m_P}\left[ 2 \phi_\vo \, \tilde\theta  \Box   \tilde\theta-  \tilde\theta^2  \Box  \phi_\vo \right].
\end{align}
The first term leads to a decay width which is mass suppressed, that was considered in \cite{Cicoli:2012aq}. However, the second term leads to a decay width proportional to $\Gamma_0$. Thus, the decay width into open string axions in minimal LVS is given by 
\begin{align}
\Gamma    \left(\phi_\vo \rightarrow \tilde\theta \tilde\theta \right)    =    \frac{1}{16} \,\Gamma_0\,.
\end{align}
While this is not as significant as moduli decays to closed string axions for the production of dark radiation, it slightly increases the tension for minimal LVS models which consider open string axions as the QCD axion. Notice moreover that the modulus decay width into the radial part is instead mass suppressed since it arises from an interaction term of the form
\begin{equation}
\mathcal{L}\supset \frac{1}{\sqrt{6}}\,\frac{\phi_\vo}{m_P}\,\tilde\rho \Box\tilde\rho.
\end{equation}

The decay to a bulk axion and an open string axion is also possible. This interaction comes also from expanding Eq.~(\ref{eq:matterKahlerPotentialMinimalLVS}), leading to a Lagrangian of the form 
\begin{align}
\mathcal{L}    &    \supset    \frac{{\rm i}}{3}    \frac{\phi_\vo}{m_P^2} \left( C \,  \partial_\mu \overline{C} \, \partial^\mu a_b-   \overline{C} \,  \partial_\mu C \,\partial^\mu a_b \right).
\end{align}
When $\tilde\rho$ takes on its VEV, this becomes 
\begin{align}
\mathcal{L}    &    \supset \frac23\left(\frac{\langle\tilde\rho\rangle}{m_P}\right)\frac{\phi_\vo}{m_P}\, \partial_\mu \tilde\theta\, \partial^\mu   a_b
\end{align}
and although this term leads to a decay width proportional to $\Gamma_0$, it is also suppressed by $\langle \tilde\rho \rangle / m_P \ll 1$. Thus, we find the contributions from the $\phi_\vo \rightarrow \tilde\theta a_b$ decay to be negligible.

We now consider open string axions within fibred LVS. In this scenario, we consider a general K\"ahler potential for matter fields of the form
\begin{align}
\frac{K}{m_P^2} &\supset \frac{C  \overline{C}}{\left( T_1  + \overline{T}_1 \right)^{x_1}  \left(  T_2   + \overline{T}_2  \right)^{x_2} }
\end{align}
where $x_1$ and $x_2$ are constants which fix the moduli dependence of the matter K\"ahler metric. These constants are not entirely arbitrary, but rather primarily fixed by the brane configuration. 
Based on results from toroidal orientifolds - which have a similar volume scaling ($\mathcal{V} = \sqrt{\tau_1 \tau_2 \tau_3}$) to the fibred Calabi-Yau case ($\mathcal{V} = \sqrt{\tau_1} \tau_2$) through the limit $\tau_3 \rightarrow \tau_2$ - in the case at hand where the matter fields live on D3-branes, \cite{Aparicio:2008wh} suggests two cases
\begin{itemize}
\item $x_1=1$ and $x_2=0$
\item $x_1=0$ and $x_2=1$\,.
\end{itemize}
Notice that both of them reproduce the scaling of Eq.~(\ref{eq:matterKahlerPotentialMinimalLVS}).

Keeping for now the general form, we can utilize the field redefinitions from Eq.~(\ref{redefinitions}) and have the following K\"ahler potential
\begin{align}
\frac{K}{m_P^2} &\supset   \frac12 C \overline{C} \left(1 - x_1 \sqrt{\frac{2}{n_1}} \frac{\phi_1}{m_P} - x_2\sqrt{\frac{2}{n_2}} \frac{\phi_2}{m_P}\right)
\end{align}
after expanding the exponentials. The relevant interaction terms in the Lagrangian in terms of the moduli mass eigenbasis turn out to be
\bea
\mathcal{L} &\supset& \frac{m_P}{2\sqrt{6}}
\left( C \Box\overline{C} + \overline{C}\Box C \right)\phi_{\mathcal{V}} \\   
&+& \frac{m_P}{2\sqrt{6}} \left(x_2\sqrt{\frac{n_1}{n_2}} - x_1 \sqrt{\frac{n_2}{n_1}}  \right) \left( C \Box \overline{C} + \overline{C}  \Box  C \right) \phi_u \nonumber
\eea
where we have explicitly used the constraints $n_1 + n_2 = 3$ and $x_1+x_2=1$. Focusing on the transverse mode, $\phi_u$, we can write down the interaction with the open string axion once the radial component acquires a VEV
\be
\mathcal{L}    \supset    \frac{1}{2\sqrt{6}\,m_P}  \left(x_2 \sqrt{  \frac{n_1}{n_2} } - x_1  \sqrt{  \frac{n_2}{n_1}}    \right)   \left[ 2 \phi_u  \, \tilde\theta  \Box\tilde\theta  -\tilde\theta^2 \Box \phi_u \right]. \nonumber
\ee
Once again, the normalization of the kinetic terms gives $\langle \tilde\rho \rangle = f_{\theta}$. Specializing to $n_1 = 1$ and $n_2 = 2$, which reproduces the volume form (\ref{voform}), the decay to open string axions is given by 
\begin{align}
\Gamma  \left( \phi_u \rightarrow  \tilde\theta \tilde\theta  \right) &=  \begin{cases}
\Gamma_0/8 & \text{for }\, x_1 = 1 \text{ and }x_2=0 \\
\Gamma_0/32 & \text{for } x_1 = 0 \text{ and }x_2=1\,.
\end{cases}
\end{align}
Thus, we see an enhancement compared to the minimal LVS case if the K\"ahler metric depends on $\tau_1$, and a reduction if the K\"ahler metric depends on $\tau_2$. In this work, we focus on the case where $x_1 = 0$ and $x_2 = 1$, so that additional contributions to dark radiation are minimized.

\subsection{Decays to Higgses}

We analyze the Giudice-Masiero terms for separable moduli K\"ahler potentials by starting with the form 
\begin{alignat}{2}
\label{eq:gmKahlerPotentialGeneral}
\frac{K}{m_P^2} & \supset    &&    \frac{ H_u \overline{H}_u}{(T_1 + \overline{T}_1 )^{y_1} (   T_2 + \overline{T}_2 )^{y_2}} + \frac{ H_d \overline{H}_d }{( T_1 +\overline{T}_1 )^{w_1} (T_2 + \overline{T}_2  )^{w_2} }\nonumber    \\ 
& && + \frac{  Z H_u H_d   +  \text{h.c.} }{(T_1 +  \overline{T}_1  )^{k_1}( T_2 + \overline{T}_2 )^{k_2}}
\end{alignat}
where we assume $Z$ is constant with respect to the $T_i$. Moreover, making again an analogy with the toroidal case and following \cite{Aparicio:2008wh}, we set $y_1+y_2=w_1+w_2=1$. We also focus on cases where the Giudice-Masiero term has a K\"ahler metric of product form (i.e. assume that $K_{H_u H_d} = \sqrt{ K_{H_u} K_{H_d} }$), so that $k_i=(y_i+w_i)/2$ $\forall i=1,2$. This leads to three possibilities
\begin{itemize}
\item $k_1=1$ and $k_2=0$
\item $k_1=0$ and $k_2=1$
\item $k_1=k_2=1/2$\,.
\end{itemize}
We may now utilize the field redefinitions from Eq.~(\ref{redefinitions}). After expanding the exponentials and going to canonically normalized Higgs fields defined by 
\bea
\tilde{H}_u &=& 
\frac{H_u}{\sqrt{\langle T_1 + \overline{T}_1\rangle^{y_1} \langle T_2 + \overline{T}_2\rangle^{y_2}}} \nonumber \\
\tilde{H}_d &=& \frac{H_d}{\sqrt{\langle T_1 + \overline{T}_1\rangle^{w_1} \langle T_2 + \overline{T}_2\rangle^{w_2}}}\,,
\eea
we obtain the following interaction terms
\begin{alignat}{2}
K & \supset && -\tilde{H}_u \overline{\tilde{H}}_u \left(y_1  \sqrt{ \frac{2}{n_1} } \frac{\phi_1}{m_P}  + y_2  \sqrt{ \frac{2}{n_2} }  \frac{\phi_2}{m_P} \right) \nonumber     \\
& &&  - \tilde{H}_d \overline{\tilde{H}}_d\left(w_1 \sqrt{\frac{2}{n_1} }\frac{\phi_1}{m_P} +  w_2\sqrt{ \frac{2}{n_2}}\frac{\phi_2}{m_P}\right) \\ &
&&- ( Z \tilde{H}_u \tilde{H}_d + \text{h.c.} ) \left(k_1  \sqrt{\frac{2}{n_1}}  \frac{\phi_1}{m_P}+  k_2\sqrt{ \frac{2}{n_2}}  \frac{\phi_2}{m_P}    \right) \nonumber
\end{alignat}
which generate the following contributions to the interaction Lagrangian
\begin{alignat}{2}
\mathcal{L} &\supset    &&
\left( \tilde{H}_u  \Box\overline{\tilde{H}}_u  + \overline{\tilde{H}}_u  \Box \tilde{H}_u\right) \left(\frac{y_1}{\sqrt{2n_1}}\frac{\phi_1}{m_P}  + \frac{y_2}{\sqrt{2n_2}} \frac{\phi_2}{m_P} \right)
\nonumber    \\ &    &&+\left( \tilde{H}_d \Box\overline{\tilde{H}}_d + \overline{\tilde{H}}_d  \Box \tilde{H}_d \right)\left( \frac{w_1}{\sqrt{2n_1}}\frac{\phi_1}{m_P} +
\frac{w_2}{\sqrt{2n_2}} \frac{\phi_2}{m_P} \right)\nonumber    \\ & &&  + \frac{1}{m_P}\left(Z \tilde{H}_u \tilde{H}_d + \text{h.c.}\right)
\left( \frac{k_1}{\sqrt{2n_1}}  \Box  \phi_1
+ \frac{k_2}{\sqrt{2n_2}} \Box  \phi_2    \right).
\end{alignat}
We now move to the moduli mass eigenbasis using Eq.~(\ref{massEigenbasisTransformation})
\begin{alignat}{2}
\label{gmTermsMassEigenbasis}
\mathcal{L} & \supset    && \frac{1}{\sqrt{6}} \left(  
\tilde{H}_u \Box \overline{\tilde{H}}_u  + \overline{\tilde{H}}_u \Box \tilde{H}_u \right)\frac{\phi_\vo}{m_P}    \nonumber    \\ &
&&+ \frac{1}{\sqrt{6}} \left(  \tilde{H}_d  \Box \overline{\tilde{H}}_d+ \overline{\tilde{H}}_d\Box \tilde{H}_d \right) \frac{\phi_\vo}{m_P} \nonumber    \\ & &&+  \frac{1}{\sqrt{6}\, m_P} (Z \tilde{H}_u \tilde{H}_d +\text{h.c.})\,\Box\phi_\vo \nonumber    \\ & &&+ \frac{\alpha}{\sqrt{6}}\left(\tilde{H}_u \Box\overline{\tilde{H}}_u+ \overline{\tilde{H}}_u \Box \tilde{H}_u \right)\frac{\phi_u}{m_P} \nonumber    \\ & &&+ \frac{\beta}{\sqrt{6}}\left( \tilde{H}_d \Box \overline{\tilde{H}}_d + \overline{\tilde{H}}_d  \Box \tilde{H}_d \right)\frac{\phi_u}{m_P} \nonumber     \\ & &&+ \frac{\gamma}{\sqrt{6}\, m_P}( Z \tilde{H}_u \tilde{H}_d + \text{h.c.}) \,\Box\phi_u
\end{alignat}
where we have explicitly restored the constraints $y_1+y_2=w_1+w_2=k_1+k_2=1$ and $n_1 + n_2 = 3$ and we made the definitions
\begin{align}
\alpha &\equiv  \left(y_2 \sqrt{\frac{n_1}{n_2}}   -  y_1 \sqrt{ \frac{n_2}{n_1} }  \right)\nonumber    \\
\beta & \equiv \left(w_2 \sqrt{\frac{n_1}{n_2}}  -  w_1\sqrt{\frac{n_2}{n_1}} \right)\nonumber    \\
\gamma    &  \equiv  \frac12(\alpha+\beta)=\left(k_2 \sqrt{\frac{n_1}{n_2}}   - k_1 \sqrt{\frac{n_2}{n_1}}\right).
\end{align}
Focusing on the transverse mode, $\phi_u$, we note that the final term will be dominant - the others will be suppressed by a factor of $m_H^2 / m_u^2$. The dominant term is therefore proportional to the coefficient $\gamma$ which for $n_1=1$ and $n_2=2$, reproducing Eq.~(\ref{voform}), reduces to
\be
\gamma = \frac{1}{\sqrt{2}}\left(k_2-2k_1\right).
\ee
It is worth stressing that this coefficients vanishes only for $k_2=2 k_1$ which is however never the case for our options since it would be inconsistent with the results of \cite{Aparicio:2008wh}. The fact that $\gamma\neq 0$ is crucial to avoid dark radiation overproduction, contrary to the results of \cite{Angus:2014bia} which considered $k_1 = 1/3$ and $k_2 = 2/3$. The intuitive reason why $k_2=2 k_1$ would imply no interaction between $\phi_u$ and the Higgses (at least at leading order) is that $\phi_u$ is orthogonal to the volume mode and in this case the K\"ahler moduli dependence of the Giudice-Masiero term in Eq.~(\ref{eq:gmKahlerPotentialGeneral}) would scale exactly as an inverse power of the overall volume since $\tau_1^{-k_1}\tau_2^{-k_2} = \vo^{-2k_1}$ for $k_2=2k_1$. 

Let us close this section by commenting briefly on a recent result. In \cite{Hebecker:2022fcx}, it was argued that there should be an additional coupling between the Higgs sector and the moduli. This coupling is induced through loop corrections to the Higgs mass term due to the volume dependence of the running from some high scale (e.g. the Kaluza-Klein scale) down to the SUSY breaking scale, and is of the form 
\begin{align}
\mathcal{L} &\supset c_{\rm loop}\left(\frac{m_u}{m_\vo}\right)^2\,  \frac{m_0^2}{m_P}\, \phi_u h^2
\end{align}
where $h$ is the SM Higgs field, $c_{\rm loop} \sim 1/(16\pi^2)$ is a typical loop coefficient and the $(m_u/m_\vo)^2\sim \vo^{-1/3}$ factor is due to the mixing between $\phi_\vo$ and $\phi_u$ induced by string loop corrections to $K$ \cite{Cicoli:2012cy}. Thus, we would expect the decay width to be proportional to 
\be
\Gamma \sim c_{\rm loop}^2 \left(\frac{m_0}{m_\vo}  \right)^4 \frac{m_u^3}{m_P^2}.
\label{Gamma}
\ee
In the case of D3-branes at singularities, we have seen that scalar masses scale as $m_0\sim M_{1/2}\sim m_\vo/\sqrt{\vo}$ in the ultralocal limit, and so we can safely neglect this induced coupling. On the other hand, in the local limit $m_0\simeq M_{1/2}\sqrt{\vo}\simeq m_\vo$, and so a more careful study is needed. The exact ratio between SUSY scalar masses and the mass of the volume mode has been derived in \cite{Cicoli:2015bpq} and takes the form
\be
\left(\frac{m_0}{m_\vo}\right)^4\simeq \left(\ln \vo\right)^2.
\ee
At first sight this factor might look large but it has to be multiplied by $c_{\rm loop}^2\sim 1/(16\pi^2)^2$ in the prefactor in Eq.~(\ref{Gamma}). For the values of the volume which we will consider in this paper, $\vo\lesssim\mathcal{O}(10^7)$ (that also keep soft terms above the TeV-scale), one has $c_{\rm loop}^2 \left(m_0/m_\vo \right)^4 \lesssim\mathcal{O}(0.01)$, showing that this decay channel can be neglected also in the local limit.

\subsection{Dark radiation predictions}
\label{DRpred}

Let us now derive the prediction for the production of axionic dark radiation from the decay of the lightest modulus $\phi_u$. The produced dark radiation can be parameterized by the effective number of neutrinos, $\Delta N_{\text{eff}}$, 
given by \cite{Cicoli:2012aq}
\begin{align}
\Delta N_{\text{eff}} &=3 \frac{        \rho_{\text{hid}}}{\rho_{\text{neutrinos}}}
= \frac{43}{7} \frac{ \rho_{\text{hid}}     }{\rho_{\text{SM}}}    \nonumber    \\
&=    \frac{43}{7} \frac{f_{\text{hid}}}{
1 - f_{\text{hid}}}  \left( \frac{            g_*(T_{\text{dec}})}{g_*(T_{\rm rh})} \right)^{1/3},
\end{align}
where $g_*(T)$ is the number of relativistic degrees of freedom at a given temperature $T$, while $f_{\rm hid}$ is the branching fraction into hidden sector particles (bulk axions) defined as
\be
f_{\text{hid}}    \equiv 
\frac{\Gamma( \phi_u \to a a ) }{\Gamma( \phi_u \to a a ) + \Gamma(\phi_u \to H_u H_d ) }\,.
\ee
The decay width into Higgses however depends on the SUSY breaking scale, and so we study the ultralocal and local limits separately.

\paragraph*{\textbf{Ultralocal limit:}}

In the ultralocal limit, as can be seen from Eq.~(\ref{limits}), the soft scalar masses scale as $m_0 \simeq M_{1/2}\sim m_P/\vo^2$. This mass scale is lower than the mass of $\phi_u$ since $m_0/m_u \simeq \vo^{-1/3}\ll 1$, implying that all MSSM Higgs degrees of freedom are accessible in the decay of $\phi_u$. Thus, taking the dominant decay term of Eq.~(\ref{gmTermsMassEigenbasis}), we can approximate the decay width into the Higgs sector as 
\begin{align}
\label{eq:higgsDecaysUltralocal}
\Gamma( \phi_u \rightarrow H_u H_d )  &    =
\frac{2 \gamma^2 Z^2}{48 \pi}  \frac{m_u^3}{m_P^2} \equiv c_{\text{vis}}\, \Gamma_0
\end{align}
where we defined 
\begin{align}
c_{\text{vis}} & \equiv 2 Z^2\gamma^2
\end{align}
as the approximate coefficient of visible sector decays. Thus the branching fraction into hidden sector degrees of freedom takes the form
\be
f_{\text{hid}} \simeq  \left[1+2Z^2\,\frac{\left(k_2 n_1-k_1 n_2\right)^2}{n_1^2+n_2^2}\right]^{-1}.
\ee

\begin{figure}[h]
\centering
\includegraphics[scale=0.365]{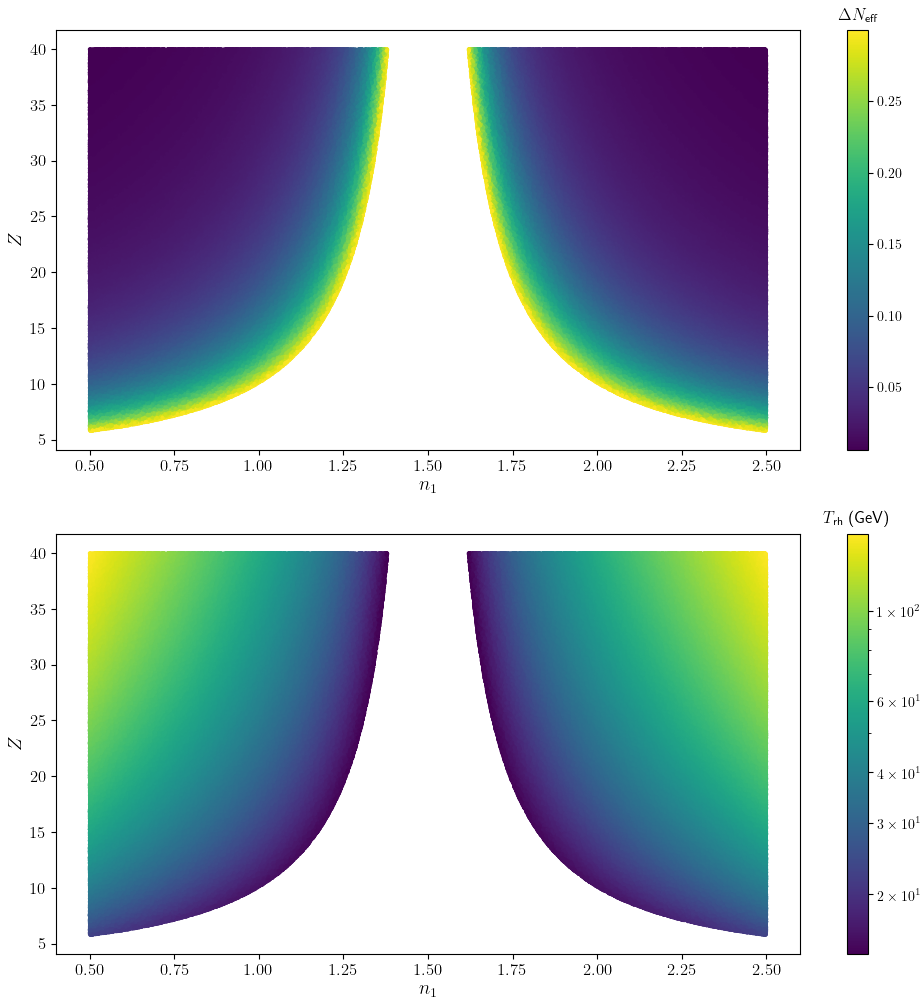}
\caption{The upper plot shows allowed $\Delta N_{\text{eff}}$ for the case $k_1 = k_2 = 1/2$ in the ultralocal limit. The lower plot shows the corresponding reheat temperature in GeV for the benchmark mass in Table \ref{tab:benchmark1A}.}
\label{fig:ultralocal_c1eq05_c2eq05}
\end{figure}

\begin{figure}[h]
\centering
\includegraphics[scale=0.364]{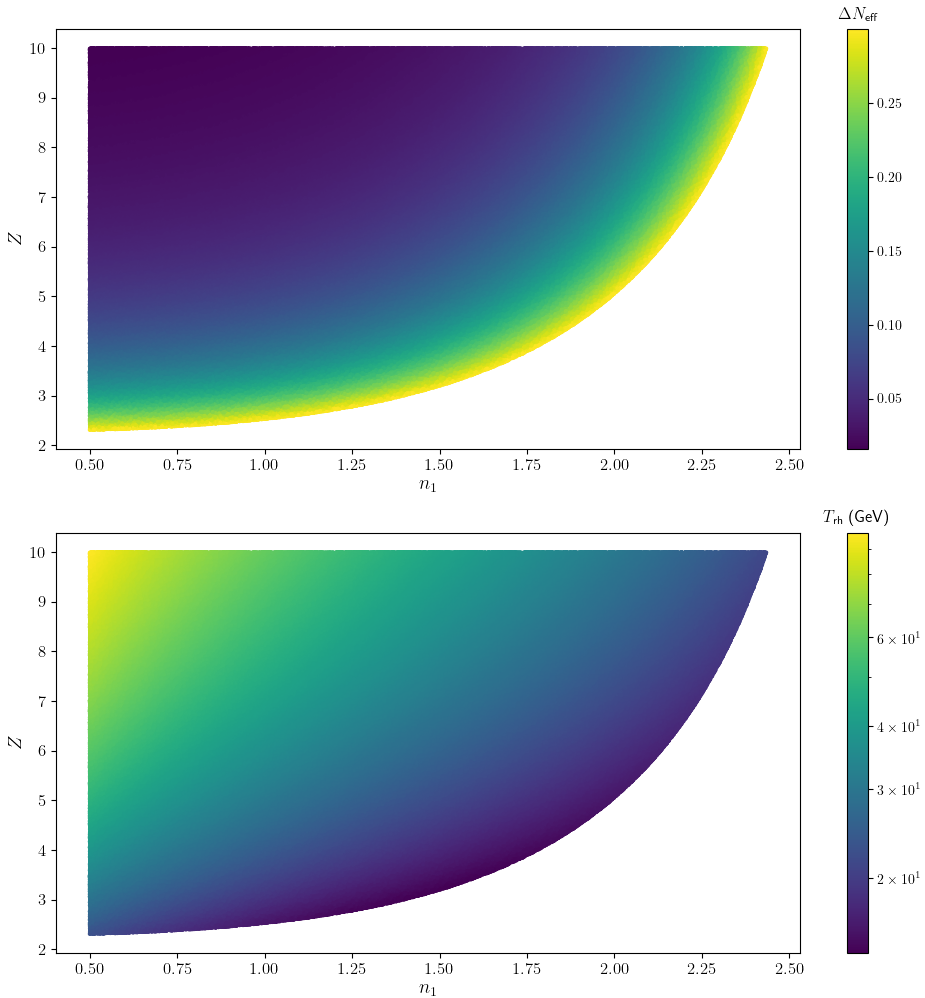}
\caption{The upper plot shows allowed $\Delta N_{\text{eff}}$  for the case $k_1 = 1$ and $k_2 = 0$ in the ultralocal limit. The lower plot again shows the corresponding reheat temperature in GeV for the benchmark mass in Table \ref{tab:benchmark1A}. This plot also describes the case $k_1 = 0$ and $k_2 = 1$ under the interchange $n_1 \leftrightarrow n_2$.}
\label{fig:ultralocal_c1eq1_c2eq0}
\end{figure}

In the upper plots of Fig.~\ref{fig:ultralocal_c1eq05_c2eq05} and Fig.~\ref{fig:ultralocal_c1eq1_c2eq0}, we display the produced dark radiation in the ($n_1$, $Z$) plane for the three different choices of $k_1$ and $k_2$ (where we also take $n_2 = 3- n_1$). Blank regions are in excess of the current $2\sigma$ bound from Planck 2018 data \cite{Planck:2018vyg}. For $n_1=1$, the best case scenario turns out to be $k_1=1$ and $k_2=0$ which requires $Z\gtrsim 2.5$ (while $k_1=0$ and $k_2=1$ would require $Z\gtrsim 5$, and  $k_1=k_2=1/2$ $Z\gtrsim 10$). Again this result can be understood intuitively by recalling that $\phi_u$ is the mode orthogonal to the volume, corresponding to the ratio of K\"ahler moduli $u=\tau_1/\tau_2$, and rewriting the moduli dependence of the Giudice-Masiero term in Eq.~(\ref{eq:gmKahlerPotentialGeneral}) as a function of $\vo$ and $u$. It is then easy to check that the combination of $k_1$ and $k_2$ which gives the largest power of $u$ ($u^{2/3}$), and so the strongest coupling of $\phi_u$ to Higgses, is $k_1=1$ and $k_2=0$.

The associated reheating temperature becomes 
\begin{alignat}{2}
T_{\text{rh}}    &    \simeq     &&
\left( \frac{40 c_{\text{vis}}  ( c_{\text{vis}} + c_{\text{hid}})}{\pi^2  g_*(T_{\text{rh}}) } \right)^{1/4} \sqrt{\Gamma_0 m_P} \nonumber  \\ &=    &&    \left( \frac{80 Z^2 }{ \pi^2 g_*(T_{\text{rh}}) }
\right)^{1/4}  \left|\frac{k_2}{n_2}  - \frac{k_1}{n_1} \right|^{1/2} \sqrt{\Gamma_0 m_P}    \nonumber    \\ 
&    &&  \times \left[ 2Z^2 (k_2 n_1-k_1 n_2)^2  + n_1^2 +n_2^2 \right]^{1/4}.
\end{alignat}
As a benchmark example, we take $g_*(T_{\text{rh}}) \simeq 86.25$ and a typical value of the mass of $\phi_u$ for K\"ahler inflation models \cite{Conlon:2005jm} which we will study in Section \ref{case1a} obtaining $m_u \simeq 2.5 \times 10^7$ GeV, as can be seen from  Table \ref{tab:benchmark1A}. The corresponding reheat temperature $T_{\text{rh}}$ in GeV is shown in the lower plots of Fig.~\ref{fig:ultralocal_c1eq05_c2eq05} and Fig.~\ref{fig:ultralocal_c1eq1_c2eq0} in the ($n_1$, $Z$) plane. Blank regions again correspond to values which produce dark radiation above the current bounds.
As $f_{\text{hid}}$ is independent of $m_u$ while $T_{\text{rh}}$ depends on $m_u$ only through $\Gamma_0$, we expect these plots to qualitatively describe a wide range of moduli masses.

\paragraph*{\textbf{Local limit:}} 

In the the local limit, as can be seen from Eq.~(\ref{limits}), soft scalar masses scale as $m_0 \simeq M_{1/2}\sqrt{\vo} \sim m_P/\vo^{3/2}$. Contrary to the ultralocal limit, in this case $m_0$ is above the mass of $\phi_u$ since $m_0/m_u \simeq \vo^{1/6}> 1$, implying that some of the Higgs degrees of freedom are kinematically inaccessible for the light modulus $\phi_u$. We must therefore consider only the decays into the light degrees of freedom, involving the SM Higgs and the three would-be Goldstone bosons. Following \cite{Cicoli:2015}, we find the final term of Eq.~(\ref{gmTermsMassEigenbasis}) becomes
\begin{align}
\mathcal{L}  &\simeq- \frac{\gamma Z}{2\sqrt{6}\,m_P}
( \,| G^+ |^2 - \,(h^0)^2+\, (G^0)^2)\Box \phi_u
\end{align}
where $G^0$ and $G^+$ are the Goldstone bosons.
The total decay width into light Higgs degrees of freedom is then given by 
\be
\Gamma(\phi_u \to H_u H_d ) = \frac{1}{48 \pi} \frac{5 Z^2\gamma^2}{32}\frac{m_u^3}{m_P^2}  \equiv c_{\text{vis}} \Gamma_0
\label{eq:higgsDecaysLocal}
\ee
where we defined 
\be
c_{\text{vis}}  \equiv \frac{5 Z^2\gamma^2}{32}. \ee
We see that the reheat temperature and the branching fraction into hidden sector degrees of freedom can then be obtained from the ultralocal limit expressions, with the replacement $2Z^2 \rightarrow 5 Z^2 / 32$. The minimum value of $Z$ allowed by dark radiation constraints thus increases by a factor of about 3.6 in comparison to the ultralocal limit. With this scaling of $Z$, Fig.~\ref{fig:ultralocal_c1eq05_c2eq05} and Fig.~\ref{fig:ultralocal_c1eq1_c2eq0} describe also the local limit. Hence in the local limit the best case scenario, corresponding to $k_1=1$ and $k_2=0$, requires $Z\gtrsim 9$. In terms of tuning needed to avoid dark radiation overproduction, the ultralocal limit seems therefore to be favored. 

It is important to stress that the requirement to satisfy current observational bounds on $\Delta N_{\rm eff}$ without relying on unnaturally large values of $Z$, constrains the form of the K\"ahler metric for matter fields both at leading order, selecting $k_1=1$ and $k_2=0$, and at subleading order by focusing on the ultralocal limit which needs a cancellation of the $\vo$-dependence of the physical Yukawas at all orders.

\section{String inflation and dark matter}

In this section we explore different options for the origin of dark matter depending on the value of $\vo$ which, as can be seen from Eqs.~(\ref{eq:massSpectrum}) and (\ref{M12}), sets all the relevant mass scales, in particular the mass of the lightest modulus and the SUSY breaking scale. In a given string inflation model, $\vo$ is in general fixed by matching the observed amplitude of density perturbations $A_s$. We thus focus on two different string models where inflation is driven by a K\"ahler modulus: ($i$) K\"ahler moduli inflation \cite{Conlon:2005jm} which is a small-field model characterized by $\vo\sim 10^7$ and a Hubble scale during inflation of order $H_I\sim 5\times 10^8$ GeV; and ($ii$) Fibre inflation \cite{Cicoli:2008gp} that is a large-field model featuring $\vo\sim 10^3$ and $H_I\sim 5\times 10^{13}$ GeV.

\subsection{K\"ahler moduli inflation}
\label{case1a}

In K\"ahler moduli inflation the role of the inflaton is played by a blow-up mode which is displaced away from the minimum where its non-perturbative potential becomes exponentially flat. We shall therefore focus on the volume form (\ref{voform}) and consider $\tau_3$ as the inflaton field. Explicit realizations of K\"ahler moduli inflation require actually the existence of an additional blow-up mode which we will however ignore since it acts just as a heavy spectator field which guarantees that $\vo$ remains approximately constant during inflation. This is a small-field inflationary model since it predicts a very small tensor-to-scalar ratio at CMB horizon scales, $r\simeq 10^{-10}$. The observed amplitude of CMB scalar fluctuations fixes $\vo\sim 10^7$ corresponding to a relatively low Hubble scale during inflation, $H_I\simeq m_\vo \sim 5\times 10^8$ GeV. Interestingly, in sequestered D3-brane models, this value of $\vo$ can also correlate with TeV-scale soft terms for a mild tuning of background fluxes. In Table \ref{tab:benchmark1A} we show a benchmark example motivated by the results of \cite{Allahverdi:2020uax}. Note that no moduli suffer from the cosmological moduli problem which requires $m_{\rm moduli} \gtrsim 7.4 \times 10^4 \text{ GeV}$.

\begin{table}[h]
\begin{tabular}{||c|c||}
\hline
$W_0$ & 40 \\
\hline
$\mathcal{V}$ & $10^7$ \\
\hline
$g_s$ & $0.1$ \\
\hline
$\lambda$ & $2$ \\
\hline
$m_{3/2}$ & $6.1 \times 10^{11}$ GeV \\ 
\hline 
$M_{1/2}$ & $1.7 \times 10^3$ GeV \\
\hline
$m_{\tau_3}$ & $7.5 \times 10^{12}$ GeV \\
\hline
$m_\vo$ & $3.1 \times 10^{8}$ GeV \\
\hline
$m_u$ & $2.5 \times 10^{7}$ GeV \\
\hline
\end{tabular}
\caption{Benchmark spectrum for K\"ahler moduli inflation. The gaugino mass $M_{1/2}$ is obtained from (\ref{M12}) with $\omega = 0.01$ and $\tau_3 =1.5$.}
\label{tab:benchmark1A}
\end{table}

During inflation, the bulk moduli $\vo$ and $u$ are displaced from their minima due to inflationary dynamics \cite{Cicoli:2016olq} and are approximately frozen due to Hubble friction. At the end of inflation, as shown in \cite{Barnaby:2009wr}, strong preheating effects lead to a violent production of inflaton quanta which eventually decay to either hidden sector gauge bosons on D7-branes wrapping the inflaton 4-cycle \cite{Allahverdi:2020uax} or to lighter moduli and axions \cite{Hebecker:2022fcx}. Later on, the bulk moduli $\phi_\vo$ and $\phi_u$ come to dominate the energy density of the universe, and their decay dilutes any previous relic. On the other hand, as can be seen from (\ref{eq:massSpectrum}), the two bulk axions, $a_1$ and $a_2$, are in practice massless for $\vo\simeq 10^7$, and so cannot behave as fuzzy dark matter. They can however be dark radiation produced from the decay of the lightest modulus $\phi_u$ which determines the final reheating. The decay width into bulk axions has been derived in Eq.~(\ref{eq:alpDecayWidth}) and for $n_1$ and $n_2$ reduces to 
\be
\Gamma(\phi_u \rightarrow a a) = \frac52 \,\Gamma_0.
\label{Gammahid}
\ee
Using the results of Section \ref{DRpred}, we focus on the best case scenario where $k_1=1$ and $k_2=0$, so that the coupling of $\phi_u$ to Higgses is maximized to avoid dark radiation overproduction. We now analyse the ultralocal and local limits separately.

\subsubsection*{Ultralocal limit}

The decay width for $\phi_u$ into Higgses in the ultralocal limit is given by Eq.~(\ref{eq:higgsDecaysUltralocal}) which for $k_1 = 1$ and $k_2 = 0$ becomes
\be
\Gamma(\phi_u \rightarrow H_u H_d) = 4 Z^2\,\Gamma_0\,.
\label{Gammavis1}
\ee
The relative fraction to hidden sector radiation is then given by 
\be
f_{\text{hid}} \simeq \left(1 + \frac{8}{5} Z^2\right)^{-1}
\label{eq:fhiddenUltraLocalVariable}
\ee
while the reheat temperature is
\begin{align}
\label{eq:reheatUltraLocalVariable}
T_{\text{rh}} &=    2    \sqrt{5} \left( \frac{ Z^2  \left[ 1 + \frac{8}{5} Z^2 \right] }{ \pi^2 g_*(T_{\text{rh}}) } \right)^{1/4}
\sqrt{\Gamma_0 m_P}.
\end{align}
Taking $g_*(T_{\text{rh}}) \simeq 86.25$ and $Z = 3$ (which brings $\Delta N_{\rm eff}$ within current observational bounds), this gives
\be
T_{\text{rh}} =     18.7\text{ GeV}    \times 
\left( \frac{m_u}{2.5 \times 10^7 \text{GeV}} \right)^{3/2}.
\label{Trh}
\ee

\paragraph*{\textbf{Branching scenario:}}

Assuming the branching scenario occurs if $T_{\text{rh}} \lesssim 70 \text{ MeV}$ \cite{Allahverdi:2014ppa,Allahverdi:2013noa}, we see from Eq.~(\ref{Trh}) that it would require $m_u \lesssim 6.0 \times 10^5$ GeV, corresponding to a volume of order $\vo \gtrsim 5 \times 10^8$ (keeping the other benchmark values of Table \ref{tab:benchmark1A} fixed) that in K\"ahler moduli inflation would yield an amplitude of scalar fluctuations below the observed value. Hence the branching scenario is not viable. However, if instead one were to extend the inflationary model to include additional fields responsible to generate the observed value of $A_s$ in a way compatible with $\vo \gtrsim 5 \times 10^8$, the branching scenario might be potentially viable. Here, we assume this to be the case and provide a brief analysis of the branching scenario.

The required modulus abundance for $T_{\text{rh}} \lesssim 70 \text{ MeV}$ and $\vo \gtrsim 5 \times 10^8$ can be estimated as
\begin{align}
Y_{\phi_u} \equiv\frac{3 T_{\text{rh}}}{4 m_u}
\lesssim    8.7    \times 10^{-8}\,.
\end{align}
Setting the DM abundance equal to the observed value, i.e. taking $Y_{\text{DM}} = 5 \times 10^{-10} (1 \text{ GeV} / m_{\text{DM}})$, and recalling Eq. (\ref{YDMbranching}) for the prediction of the final DM abundance in the branching scenario, we note that smaller values of $m_{\text{DM}}$ require larger values of $m_u$ to match the observed DM abundance for a given branching ratio. However, as $m_{\text{DM}}$ decreases, the freeze-out temperature $T_{\text{f}}$ decreases, while increasing $m_u$ increases the reheat temperature $T_{\text{rh}}$. It is then imperative to check that $T_{\text{rh}} \lesssim T_{\text{f}}$, as required to have a non-thermal abundance.

\begin{figure}[h]
\centering
\includegraphics[scale=0.63]{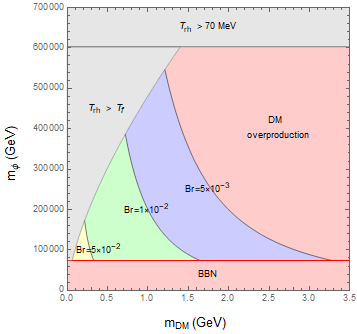}
\caption{Constraints on the branching scenario in the ultralocal limit. Regions in red are excluded, while gray regions cannot be accommodated in the branching scenario.}
\label{fig:branchingUltralocal}
\end{figure}

In Fig.~\ref{fig:branchingUltralocal}, we plot contours of the DM branching ratio which reproduce the measured abundance for given modulus and DM masses. The lower bound on $m_u$ is set by BBN constraints ($T_{\text{rh}} \gtrsim 3 \text{ MeV}$), while the upper bound on $m_u$ is set by distinguishing the thermal and non-thermal cases. We adopt $\text{Br}_{\text{DM}} \gtrsim 5 \times 10^{-3}$ as a lower bound on the branching ratio to dark matter.

Thus we see that the branching scenario in K\"ahler moduli inflation with a non-standard mechanism for the generation of the density perturbations is only potentially viable for DM masses between $100 \text{ MeV} \lesssim m_{\text{DM}} \lesssim 3.3 \text{ GeV}$ and $80 \text{ TeV} \lesssim m_u \lesssim 550 \text{ TeV}$.


\paragraph*{\textbf{Annihilation scenario:}}

\begin{figure}[h]
\centering
\includegraphics[scale=0.63]{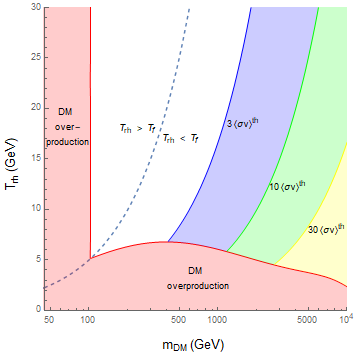}
\caption{Constraints on the annihilation scenario due to Fermi-LAT data \cite{Fermi-LAT:2016uux}. Regions in red are excluded, with the dashed line separating the thermal and non-thermal cases. Other shaded regions give DM overproduction if $\langle \sigma_{\text{ann}} v \rangle_{\text{f}}
\leq  n \langle \sigma_{\text{ann}} v \rangle_{\text{f}}^{\text{th}}$ for $n \in \{ 3, 10, 30 \}$, but may still be allowed for larger annihilation cross sections allowed by data.}
\label{fig:annihilation}
\end{figure}

In the annihilation scenario we expect an enhancement to the DM abundance relative to the thermal case. To analyze this scenario, we fit the upper bound of the annihilation cross section in the $b\bar{b}$ channel to data from \cite{Fermi-LAT:2016uux}.
If we require Eq.~(\ref{eq:annihilationDMAbundance}) to match the DM abundance, since the annihilation scenario requires $\langle \sigma_{\text{ann}} v \rangle_{\text{f}} \geq \langle \sigma_{\text{ann}} v \rangle_{\text{f}}^{\text{th}}$, we have an absolute lower bound on $T_{\text{rh}}$ from DM overproduction in the case that the annihilation cross section matches the thermal value. The constraints due to DM overproduction bounds are displayed in Fig. \ref{fig:annihilation}, where the red shaded regions are excluded. We also show potentially excluded regions depending on the annihilation cross section for values $3 \langle \sigma_{\text{ann}} v \rangle_{\text{f}}^{\text{th}}$ (blue region), $10 \langle \sigma_{\text{ann}} v \rangle_{\text{f}}^{\text{th}}$ (green region), and $30 \langle \sigma_{\text{ann}} v \rangle_{\text{f}}^{\text{th}}$ (yellow region).  These regions are excluded by DM overproduction if the annihilation cross section is between the thermal value and the given reference value, i.e. excluded if
$\langle \sigma_{\text{ann}} v \rangle_{\text{f}}
\leq  n \langle \sigma_{\text{ann}} v \rangle_{\text{f}}^{\text{th}}$ for $n \in \{ 3, 10, 30 \}$, but may still be allowed for larger cross sections (so long as they are consistent with upper bounds from data). In the thermal case, the white region may also potentially overproduce DM, depending on the annihilation cross section.
Furthermore, if one were to assume some model with multi-component DM, more stringent overproduction constraints may be expected.

For our benchmark value, which from Eq.~(\ref{Trh}) gives $T_{\rm rh}\simeq 18.7$ GeV, we see that we can have the annihilation scenario only for $m_{\text{DM}} \gtrsim 400$ GeV, while we are reduced to the thermal scenario if $m_{\text{DM}} \lesssim 400$ GeV.
Furthermore, potential DM overproduction in the annihilation scenario may be slightly improved with some minor adjustments: decreasing $\lambda$ by a factor of 3 and increasing $g_s \simeq 0.15$ then increases the reheat temperature by a factor of roughly $3/2$, easing the required annihilation cross section to match the observed DM abundance by a factor of roughly $2/3$.

\subsubsection*{Local limit}

In the local limit and in the best case scenario with $k_1=1$ and $k_2=0$, the decay of $\phi_u$ into Higgs degrees of freedom is given by Eq.~(\ref{eq:higgsDecaysLocal}) and looks like 
\be
\Gamma( \phi_u \rightarrow H_u H_d )
\simeq\frac{5}{16} Z^2 \,  \Gamma_0.
\label{Gammavis2}
\ee
We can thus estimate $f_{\text{hidden}}$ as 
\be
f_{\text{hidden}}\simeq  \left(1+ \frac18 Z^2\right)^{-1}
\ee
and the reheating temperature as
\begin{align}
T_{\text{rh}}&=  2 \sqrt{5} \left(\frac{       \frac{5}{32} Z^2 \left[1 + \frac18  Z^2 \right]}{ \pi^2 g_*(T_{\text{rh}})} \right)^{1/4}\sqrt{\Gamma_0 m_P}\,.
\end{align}
Due to the reduction in the visible sector decay width compared to the ultralocal scenario, we find a value of $Z \gtrsim 10$ is required in order to avoid dark radiation overproduction. Thus, for $Z=10$ and $g_*(T_{\text{rh}}) \simeq 86.25$, the reheating temperature becomes 
\begin{align}
T_{\text{rh}}=     20.8 \text{ GeV}\times \left( \frac{m_u}{2.5 \times 10^7 \text{ GeV}} \right)^{3/2}.
\end{align}

\paragraph*{\textbf{Branching scenario:}}

In the local limit much of our analysis follows identically to the ultralocal case.
The branching scenario can occur for $m_u \lesssim 5.6 \times 10^5 $ GeV (assuming $T_{\text{rh}} \lesssim 70$ MeV), which gives us a corresponding abundance
\be
Y_{\phi_u} \equiv  \frac{3 T_{\text{rh}} }{4 m_u}  \lesssim 9.3  \times 10^{-8}
\ee

Due to the similar modulus abundance and reheat temperature compared to the ultralocal case, we find bounds that are quantitatively similar to those in Fig. \ref{fig:branchingUltralocal}, with the maximal DM mass pushed down to $3$ GeV and the maximal modulus mass pushed down to $500$ TeV.

\paragraph*{\textbf{Annihilation scenario:}}

For the annihilation scenario in the local limit, our analysis for the ultralocal limit also applies.
From Fig. \ref{fig:annihilation}, we see that we would expect the annihilation scenario only for $m_{\text{DM}} \gtrsim 430$ GeV, and the thermal scenario for $m_{\text{DM}} \lesssim 430$ GeV.

\subsection{Fibre Inflation}
\label{case2a}

Fibre inflation models feature a Calabi-Yau volume of the form (\ref{voform}) where the role of the inflaton is played by the mode $u$ orthogonal to the overall volume. This is a large-field inflationary model which predicts a tensor-to-scalar ratio at the edge of detectability, $r\simeq 0.007$. Compared to K\"ahler moduli inflation, the Hubble scale during inflation is higher, $H_I\simeq m_u \sim 5\times 10^{13}$ GeV, and value of the volume needed to match $A_s$ is smaller, $\vo\sim 10^3$. In turn, even with sequestering, the scale of SUSY breaking turns out to be relatively high, $M_{1/2}\sim 5\times 10^{10}$ GeV. Table \ref{tab:benchmark2A} shows a benchmark example with values motivated from \cite{Cicoli:2019} where these parameter choices were shown to reproduce the observed amplitude of the density perturbations.

\begin{table}[h]
\begin{tabular}{||c|c||}
\hline
$W_0$ & 15 \\
\hline
$\mathcal{V}$ & $10^3$ \\
\hline
$g_s$ & $0.1$ \\
\hline
$\lambda$ & $2$ \\
\hline
$m_{3/2}$ & $2.3 \times 10^{15}$ GeV \\
\hline
$M_{1/2}$ & $6.3 \times 10^{10}$ GeV \\
\hline
$m_{\tau_3}$ & $9.5 \times 10^{15}$ GeV \\
\hline
$m_\vo $ & $2.0 \times 10^{14}$ GeV \\
\hline
$m_u$ & $4.4 \times 10^{13}$ GeV \\
\hline
\end{tabular}
\caption{Benchmark spectrum for Fibre inflation. The gaugino mass $M_{1/2}$ is obtained from (\ref{M12}) with $\omega=0.01$ and $\tau_3=1.5$.}
\label{tab:benchmark2A}
\end{table}

Given that in this case the inflaton is the lightest K\"ahler modulus, there is no period of early matter domination and inflationary reheating transitions directly into a radiation dominated universe via the perturbative decay of $\phi_u$. The decay width into the two ultra-light axions $a_1$ and $a_2$ is again given by Eq.~(\ref{Gammahid}). The decay rates into visible sector fields takes again the form (\ref{Gammavis1}) and (\ref{Gammavis2}) for the ultralocal and local limits respectively. However, in this case soft masses are around the intermediate scale, $M_{1/2}\sim 10^{10}$-$10^{11}$ GeV, and so neutralino DM would be grossly overproduced in both thermal and non-thermal scenarios since naively $\langle \sigma v \rangle \propto m_{\rm DM}^{-2}$ for WIMPs. We need therefore to focus on scenarios with R-parity breaking where neutralinos are unstable. 

The origin of DM therefore has to be different. Ref. \cite{Cicoli:2018asa} argued that the potential of Fibre inflation is rich enough to be able to generate primordial black holes that can constitute all of DM in an appropriate mass range, with the associated production of detectable secondary gravity waves \cite{Cicoli:2022sih}. Given that this possibility involves a consistent tuning of the underlying parameters, we instead investigate whether DM can be made up of axions produced via the standard misalignment mechanism. These can be either the two ultra-light axions $a_1$ and $a_2$ which for $\vo\simeq 10^3$ have masses and decay constants in the right ballpark to behave as fuzzy dark matter \cite{Cicoli:2021gss}, or the QCD axion realized as the phase of a charged open string field living on the visible sector D3-brane stack \cite{Cicoli:2013cha}. However, strong constraints on the type of axion are imposed due to existing bounds on isocurvature perturbations for cases where the Peccei-Quinn symmetry is broken during inflation so that the axion is effectively massless and undergoes quantum fluctuations of order $H_I/(2\pi)$. More precisely, when an axion (which can be either the QCD axion or an ultra-light axion) saturates the DM relic abundance for a decay constant $f_a>H_I$ and an initial misalignment angle $\theta_i$, the bound on the inflationary scale from the measured amplitude of isocurvature modes is \cite{Visinelli:2009zm, Planck:2018jri}
\be
H_I \lesssim 1.4\times 10^{-5}\,f_a\,\theta_i.
\ee
For $H_I \simeq 5\times 10^{13}$ GeV, this bound can be translated into a strong bound on $f_a$ of the form
$f_a\,\theta_i \gtrsim m_P$, implying that $f_a$ has to be at least of order the Planck scale since $\theta_i \in [0,2\pi]$. As shown in \cite{Cicoli:2021gss}, each of the two bulk axions could be fuzzy dark matter with $f_{a_1}\sim f_{a_2}\sim m_p/\vo^{2/3}$ which is definitely below $m_P$ for $\vo \simeq 10^3$. This implies that fuzzy dark matter is not allowed in Fibre inflation. 

The other possibility is to look at the QCD axion. As shown in Section \ref{OpenAxions}, the QCD axion can arise from the phase of a matter field living on the D3-brane stack. Its decay constant $f_{\rm QCD}$ is set by the VEV of the associated radial field that is stabilized by an interplay of D- and F-terms. As shown in \cite{Cicoli:2013cha}, the value $f_{\rm QCD}$ depends on the level of sequestering of the visible sector from the source of SUSY breaking in the bulk. In particular, focusing on the benchmark mass spectrum of Tab. \ref{tab:benchmark2A}, we have
\bea
&&\text{local limit}: \qquad\quad f_{\rm QCD} \simeq m_{3/2} \simeq 10^{15}\,{\rm GeV} \nonumber \\
&&\text{ultralocal limit}:\quad f_{\rm QCD} \simeq M_{1/2} \simeq 5\times 10^{10}\,{\rm GeV}. \nonumber
\label{AxionSequestering}
\eea
In the local limit, $f_{\rm QCD}$ is therefore above $H_I$, implying that this case is also ruled out by isocurvature bounds. On the other hand, in the ultralocal limit, $f_{\rm QCD}< H_I$, and so during inflation the axion is heavy given that the Peccei-Quinn symmetry is not broken yet. Thus in this case the system if free from isocurvature bounds, and the decay constant of the open string axion is in the right energy regime to constitute all of dark matter \cite{Gorghetto:2020qws}. 

The ultralocal limit seems therefore favored to describe dark matter. In this case Eq.~(\ref{eq:fhiddenUltraLocalVariable}) and Eq.~(\ref{eq:reheatUltraLocalVariable}) give again the branching ratio to hidden light degrees of freedom and the reheat temperature, respectively. Taking $g_*(T_{\text{rh}}) = 106.75$ and $Z=3$ (to avoid dark radiation overproduction), we have a relatively high reheat temperature of order
\be
T_{\text{rh}}   \simeq     4.07    \times 10^{10} \text{ GeV}    \times    \left( \frac{m_u }{4.4 \times 10^{13} \text{ GeV} }    \right)^{3/2}.
\ee

\section{Conclusions}
\label{Concl}

In this work, we have studied the production of dark matter and dark radiation after reheating in string inflation models with a fibred Calabi-Yau structure and the visible sector on D3-branes.

We have imposed several physical constraints, related to the need to match observed quantities, like the amplitude of the density perturbations generated during inflation and the dark matter density, or coming from the requirement to avoid the overproduction of dark radiation and isocurvature modes. The interplay between these different physical quantities is determined by the consistency of the underlying UV model which fixes also the scale of supersymmetry breaking, that determines the WIMP mass scale, and the decay constant of both closed and open string axions. We have seen that this UV correlation is very powerful in constraining the nature of dark matter. 

In fact, choosing benchmark points preferred by cosmological data for each of our string inflation scenarios, we have found distinct predictions for the nature of dark matter which are compatible with current experimental bounds. In the case of K\"ahler moduli inflation, we have found TeV-scale WIMPs which can easily reproduce the observed DM density in both the local and ultralocal limits. The case of Fibre inflation instead predicts open string axions as the DM candidate within the ultralocal limit, where isocurvature constraints exclude the local limit. 

While we found that an excess of dark radiation is generic in fibred LVS models, we have shown that this can be compatible with the latest Planck measurements without any significant tuning. Let us also point out that Planck bounds on $N_{\rm eff}$ can be relaxed if local measurements of $H_0$ are assumed as a prior, and a small but non-zero amount of dark radiation could also be an intriguing effect to weaken the present $H_0$ tension \cite{Brinckmann:2020bcn}. Let us finally stress that our analysis involved KMI and FI for fibred Calabi-Yaus with the MSSM on D3-branes. We will instead leave the study of the case where the MSSM is on D7-branes for future work \cite{future}.

\section*{Acknowledgments}

We would like to thank Rouzbeh Allahverdi and Arthur Hebecker for helpful discussions. The work of KS is supported in part by DOE Grant DE-SC0009956.

\appendix

\bibliography{apssamp}

\end{document}